\documentclass[sigconf,pbalance,nonacm]{acmart}

\AtBeginDocument{%
  }

\usepackage{pgfplots}
\pgfplotsset{compat=1.18}
\usetikzlibrary{positioning}
\usepackage{enumitem}

\begin{document}

\title[EGR]{EGR: Embedding-Native Generative Retrieval with a Shared LLM}

\author{Xiaodong Liu}
\affiliation{%
  \institution{Snap Inc.}
  \city{Bellevue}
  \state{WA}
  \country{USA}
}
\email{xliu9@snapchat.com}

\author{Congfei Zhang}
\affiliation{%
  \institution{Snap Inc.}
  \city{Bellevue}
  \state{WA}
  \country{USA}
}
\email{czhang3@snapchat.com}

\author{Hsiang-wei Chao}
\affiliation{%
  \institution{Snap Inc.}
  \city{Seattle}
  \state{WA}
  \country{USA}
}
\email{hchao@snapchat.com}

\author{Siman Wang}
\affiliation{%
  \institution{Snap Inc.}
  \city{Bellevue}
  \state{WA}
  \country{USA}
}
\email{swang7@snapchat.com}

\author{Tong Zhao}
\affiliation{%
  \institution{Snap Inc.}
  \city{Bellevue}
  \state{WA}
  \country{USA}
}
\email{tzhao@snapchat.com}

\author{Xiao Bai}
\affiliation{%
  \institution{Snap Inc.}
  \city{Palo Alto}
  \state{CA}
  \country{USA}
}
\email{xbai@snapchat.com}

\author{Vincent Zhang}
\affiliation{%
  \institution{Snap Inc.}
  \city{Bellevue}
  \state{WA}
  \country{USA}
}
\email{wzhang7@snapchat.com}

\author{Jingxiao Ma}
\affiliation{%
  \institution{Snap Inc.}
  \city{Bellevue}
  \state{WA}
  \country{USA}
}
\email{jma3@snapchat.com}

\author{Zhe Liu}
\affiliation{%
  \institution{Snap Inc.}
  \city{Palo Alto}
  \state{CA}
  \country{USA}
}
\email{zliu11@snapchat.com}

\author{Wenfeng Zhuo}
\affiliation{%
  \institution{Snap Inc.}
  \city{Palo Alto}
  \state{CA}
  \country{USA}
}
\email{wzhuo@snapchat.com}

\author{Zichu Li}
\affiliation{%
  \institution{Snap Inc.}
  \city{Palo Alto}
  \state{CA}
  \country{USA}
}
\email{zli12@snapchat.com}

\author{Jitin Krishnan}
\affiliation{%
  \institution{Snap Inc.}
  \city{Bellevue}
  \state{WA}
  \country{USA}
}
\email{jkrishnan2@snapchat.com}

\author{Yunzhi Zhou}
\affiliation{%
  \institution{Snap Inc.}
  \city{Palo Alto}
  \state{CA}
  \country{USA}
}
\email{yzhou10@snapchat.com}

\author{Yajun Wang}
\affiliation{%
  \institution{Snap Inc.}
  \city{Palo Alto}
  \state{CA}
  \country{USA}
}
\email{ywang30@snapchat.com}

\author{Jinchao Li}
\affiliation{%
  \institution{Snap Inc.}
  \city{Bellevue}
  \state{WA}
  \country{USA}
}
\email{jli18@snapchat.com}

\author{Yu Zhang}
\affiliation{%
  \institution{Snap Inc.}
  \city{Palo Alto}
  \state{CA}
  \country{USA}
}
\email{yzhang3@snapchat.com}

\renewcommand{\shortauthors}{Liu et al.}

\begin{abstract}
Generative retrieval is increasingly popular in large-scale recommendation and advertising systems, yet current methods introduce practical complications. Semantic-ID methods rely on quantization, mutable identifier vocabularies, and token-to-item grounding; embedding-based pipelines train the item encoder separately from the query generator, which limits user-item alignment. We propose \textbf{EGR}, an \textbf{E}mbedding-native \textbf{G}enerative \textbf{R}etrieval framework for recommendation and advertising. EGR uses a single shared LLM to learn item representations from item metadata and user representations from interaction histories in one embedding space. Items are indexed directly as dense vectors, and user histories are encoded as dense retrieval queries. Joint contrastive training groups related items and aligns queries with their target items. We evaluate EGR on public benchmarks, industrial data, and live deployment. EGR outperforms published baselines on Amazon Reviews; on Snap DPA, it scales with data, handles cold-start items, and benefits from multimodal input. In production, EGR delivers a $+2.91\%$ conversion-rate lift, simplifying system design while improving retrieval quality and ad performance.
\end{abstract}



\maketitle

\section{Introduction}
Large-scale candidate retrieval is a foundational layer in modern recommendation and advertising systems~\cite{covington2016youtube,cheng2016widedeep,ying2018pinsage,yi2019twotower}. For each request, a retriever must select a compact set of relevant items from catalogs that may contain hundreds of millions of items and change continuously, all under strict latency and reliability budgets. A useful retriever must therefore combine personalization with catalog coverage, freshness, and operational simplicity.

Industrial retrieval stacks are commonly built around two-tower embedding models and their sequential extensions~\cite{kang2018sasrec,sun2019bert4rec}. These systems encode queries and items into a shared vector space and retrieve candidates with approximate nearest-neighbor (ANN) search. Generative retrieval (GR) offers a different route: a sequence model conditions on user history and produces the next target, either as a sequence of item identifiers or as a dense retrieval representation. This paradigm has shown strong results in academic benchmarks~\cite{rajput2023tiger,zhai2024hstu} and in recent industrial deployments~\cite{agarwal2026pinrec,he2025plum}.

Existing GR methods fall into two main families. \emph{SID-based} methods~\cite{rajput2023tiger,he2025plum,deng2025onerec,ye2025das,xue2026gr4ad} map items to discrete semantic identifier sequences and train a generator to emit those sequences. They require an item encoder, a quantizer, a generator, and a grounding layer, and catalog updates require assigning identifiers to new or changed items. \emph{Embedding-based} methods~\cite{agarwal2026pinrec,liang2025tbgrecall} generate continuous vectors for ANN search and avoid discrete identifiers, but they train the item encoder separately from the query generator. In that setting, the indexed item space is fixed before the query side is trained, which can make user-item alignment a post-hoc problem.

\begin{figure}[t]
  \centering
  \includegraphics[width=0.85\linewidth]{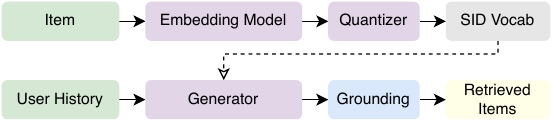}\\
  {\small (a) SID-based GR}\\[0.6em]
  \includegraphics[width=0.85\linewidth]{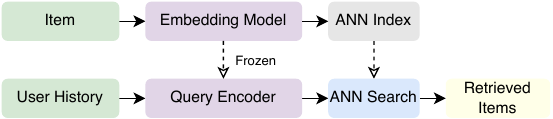}\\
  {\small (b) Embedding-based GR}\\[0.6em]
  \includegraphics[width=0.85\linewidth]{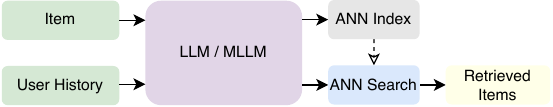}\\
  {\small (c) Embedding-Native GR (ours)}
  \caption{Comparison of generative retrieval families. (a) \textbf{SID-based GR} encodes items, quantizes them into a semantic-ID vocabulary, and trains a generator to emit SID token sequences that are grounded back to items at serve time. (b) \textbf{Embedding-based GR} uses separate item and query encoders; the item embedding model is pretrained or trained first, then fixed during query encoder training, so the indexed item embeddings are not optimized for the generative retrieval objective. (c) \textbf{Embedding-Native GR (ours)} uses one shared LLM backbone for both items and user histories, jointly trained so the indexed item space and the query side evolve together.}
  \Description{Comparison of three generative retrieval families. SID-based generative retrieval maps items into semantic identifiers and grounds generated IDs back to items. Embedding-based generative retrieval uses a separate item embedding model and query encoder; the item model is pretrained or trained first and then fixed while the query encoder is trained, so the indexed item embeddings are not optimized for the generative retrieval objective. Embedding-native generative retrieval (ours) uses a shared LLM backbone for item and history encoding and jointly trains both retrieval paths.}
  \label{fig:method-comparison}
\end{figure}

We propose \textbf{EGR}, an embedding-native generative retrieval framework (Figure~\ref{fig:method-comparison}). EGR uses one shared LLM backbone in two paths. The item path encodes item metadata into dense vectors that are written directly to an ANN index. The user-history path consumes a chronological sequence of item engagements and emits a dense query vector. As the two paths share parameters and are optimized jointly, the indexed item space and the user-conditioned query space are learned together rather than stitched together after separate training. Our contributions are:
\begin{itemize}[leftmargin=*]
  \item We introduce an embedding-native GR design that removes the quantization and grounding of SID-based methods and the separation of item embedding and user embedding in embedding-based methods, while preserving standard ANN serving.
  \item We use a shared LLM that natively aligns user and item representations in one embedding space, leveraging pretrained priors for multimodal and cold-start generalization, jointly trained with item-pair and history-to-target contrastive objectives.
  \item We evaluate EGR on public benchmarks, industrial-scale data, and a live online deployment. EGR outperforms the evaluated published baselines on Amazon Reviews. On industrial Snap DPA data, EGR (i) scales monotonically with more training data, (ii) remains strong on cold-start items, and (iii) benefits from multimodal input; in a live A/B test, it yields a $+2.91\%$ CVR lift.
\end{itemize}

\section{Related Work}
\label{sec:related-work}

\textbf{Sequential recommendation.} Sequential recommenders such as SASRec~\cite{kang2018sasrec} and BERT4Rec~\cite{sun2019bert4rec} established transformer-based next-item prediction as a strong pre-generative baseline. They use learned item-ID embeddings as model parameters rather than content-based item encoders.

\noindent \textbf{SID-based generative retrieval.} TIGER quantizes item-content embeddings and trains a transformer to predict the next item's identifier sequence~\cite{rajput2023tiger}. PLUM scales this direction to industrial YouTube recommendation with LLM adaptation and improved semantic IDs~\cite{he2025plum}. Subsequent systems refine identifier construction, alignment, and serving. DAS aligns user and ad semantic IDs through dual quantization~\cite{ye2025das}, IDGen learns identifier construction jointly with the recommendation objective~\cite{tan2024idgenrec}, OneRec unifies retrieve-and-rank through SID generation with preference alignment~\cite{deng2025onerec}, and GR4AD and GPR adapt generative recommendation to advertising with value-aware training; GR4AD additionally adds production serving optimizations and GPR proposes a one-model paradigm~\cite{xue2026gr4ad, zhang2026gpr}. Recent work also moves beyond purely numeric SIDs. For example, GRLM proposes structured term IDs drawn from the LLM's native vocabulary to reduce grounding hallucination during generation~\cite{zhang2026grlm}. A parallel line of work recognizes that item identifiers should not be learned independently of the downstream generator: PIT co-evolves item tokenization with the generative recommender for dynamic industrial streams~\cite{wang2026pit}, and the Spotify semantic-ID study shows that semantic IDs built from jointly fine-tuned bi-encoder embeddings transfer well across search and recommendation~\cite{penha2025spotifyids}. These methods differ in how they build, align, and serve identifiers, but they retain a discrete identifier layer between item representation and retrieval. EGR removes that layer and trains the retrieval embedding space directly.

\begin{figure*}[!t]
  \centering
  \includegraphics[width=\textwidth]{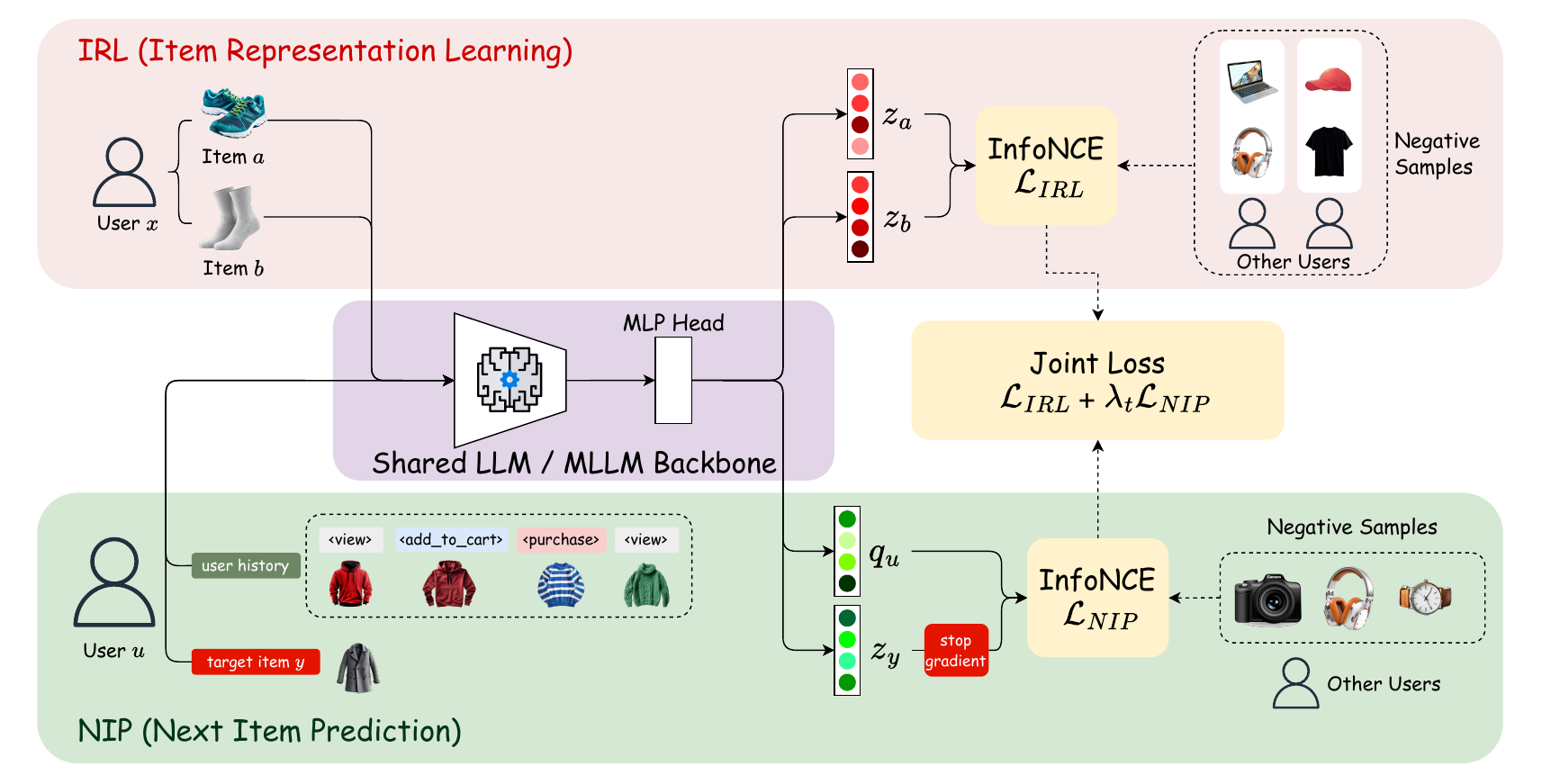}
  \caption{Overall framework of EGR. A shared LLM with a projection head encodes both item and user histories. IRL applies a symmetric InfoNCE on the item pair $(p_a, p_b)$, shaping the indexed item space. NIP applies a history-to-target InfoNCE from the user-conditioned query embedding to the target item embedding, with stop-gradient on the target. Joint training combines the two losses as $\mathcal{L}_{\mathrm{IRL}} + \lambda_t \mathcal{L}_{\mathrm{NIP}}$.}
  \Description{Framework diagram of EGR. A shared LLM backbone with a projection head encodes both items and user histories. The NIP path encodes a user history (with conversion-type tokens) and a target item into embeddings combined by an InfoNCE loss, with a stop-gradient marker on the target branch. The IRL path encodes two co-engaged items from the same user into embeddings combined by a symmetric InfoNCE loss. The two losses are summed into a joint objective L\_IRL plus lambda times L\_NIP.}
  \label{fig:egr-overview}
\end{figure*}

\noindent \textbf{Embedding-based generative retrieval.} PinRec~\cite{agarwal2026pinrec} generates dense query representations for Pinterest's Homefeed, Search, and Related Pins surfaces and retrieves through ANN, avoiding semantic-ID quantization altogether. It uses a two-model design with a separate item encoder and query generator, together with outcome-conditioned generation and temporal multi-token inference. Other systems in this family include TBGRecall~\cite{liang2025tbgrecall}, which uses a session-level representation for ANN retrieval in e-commerce. EGR differs from PinRec in two central ways. First, a single shared LLM backbone produces both item and query embeddings through the same parameters and projection head. Second, joint training updates the indexed item space and the query side together, rather than freezing the item encoder before training the query generator.

\section{Method}
\label{sec:method}

\subsection{Overall Framework}
Figure~\ref{fig:egr-overview} shows the EGR architecture. EGR learns a shared embedding space for items and user-conditioned queries. Each item is represented by its metadata, including title, brand, category, and image; each user is represented by a chronological sequence of engagement events, where each event contains a conversion type and an item. A shared LLM backbone encodes both inputs, followed by a projection head that compresses the backbone's hidden state to a lower-dimensional retrieval embedding for compact ANN serving. The item path maps item metadata to an indexed item embedding, and the user-history path maps recent engagements to a query embedding. Because both paths share parameters and emit vectors in the same space, next-item prediction becomes dense retrieval: at inference time, EGR emits a query vector and retrieves items by maximum inner product. This design is \emph{embedding-native}: it never predicts item-specific tokens, quantizes item vectors, or grounds generated identifiers back to items. The two paths are optimized jointly with the contrastive objectives described below.

\subsection{Item Representation Learning (IRL)}
\label{sec:irl}
\noindent\textbf{Item encoder.} Let each item $i$ have metadata $x_i$ (title, brand, category, and image). IRL trains item embeddings with co-engagement contrastive learning. The encoder serializes $x_i$ into a fixed prompt and passes it through the shared LLM backbone $f_\theta$ followed by a projection head $h_\phi$, producing an L2-normalized vector
\begin{equation}
  z_i = \frac{h_\phi(f_\theta(x_i))}{\|h_\phi(f_\theta(x_i))\|_2},
\end{equation}
which is inserted directly into the ANN index. Unlike semantic-ID systems, $z_i$ is not quantized and requires no token-to-item grounding; a new or updated item can enter the index as soon as its metadata is encoded.

\noindent\textbf{Pair contrastive objective.} For a consecutive co-engaged item pair $(p_a, p_b)$ from the same user's chronological sequence, IRL applies a symmetric InfoNCE loss~\cite{oord2018cpc}:
\begin{equation}
  \mathcal{L}_{\mathrm{IRL}} = \tfrac{1}{2}\bigl[\mathrm{InfoNCE}(z_a, z_b) + \mathrm{InfoNCE}(z_b, z_a)\bigr],
\end{equation}
where $\mathrm{InfoNCE}(a,b) = -\log \frac{\exp(a^\top b / \tau)}{\sum_{b' \in \{b\}\cup\mathcal{N}} \exp(a^\top b' / \tau)}$, $\mathcal{N}$ is the set of in-batch negatives, $\tau$ is a learned temperature, and all vectors are L2-normalized. IRL uses a co-engagement structure to shape the item space that will later be indexed for retrieval.

\subsection{Next Item Prediction (NIP)}
\label{sec:nip}
\noindent\textbf{History encoder.} Let user $u$ be represented by a chronological history $H_u=(e_1,\ldots,e_T)$ of (conversion-type, item) events. NIP trains the user-history encoder to produce a query that retrieves the target item. EGR serializes $H_u$ in oldest-to-newest order, prefixing each event with a conversion-type token (\texttt{<view>}, \texttt{<add\_to\_cart>}, \texttt{<purchase>}) that reflects different user intents and joins events with separators. The serialized history is encoded through the same backbone and projection head:
\begin{equation}
  q_u = \frac{h_\phi(f_\theta(H_u))}{\|h_\phi(f_\theta(H_u))\|_2}.
\end{equation}
At serving time, $q_u$ retrieves items by maximum inner product against the indexed $z_i$.

\noindent\textbf{History-to-target objective.} The NIP loss is a history-to-target InfoNCE objective with a stop-gradient on the target embedding:
\begin{equation}
  \mathcal{L}_{\mathrm{NIP}} = \mathrm{InfoNCE}\bigl(q_u,\;\mathrm{sg}(z_y)\bigr),
\end{equation}
where $y$ is the target item and $\mathrm{sg}(\cdot)$ denotes a stop-gradient. The target embedding $z_y$ is recomputed by the current backbone at every step, but the stop-gradient prevents $\mathcal{L}_{\mathrm{NIP}}$ from updating the target branch for that example. Because the backbone is shared, NIP gradients still update the model through the history branch and can influence future item embeddings through the shared parameters. The stop-gradient prevents per-example target pulls from destabilizing the indexed item space.

\subsection{Joint Training}
\label{sec:joint-training}
\noindent\textbf{Total objective.} EGR optimizes IRL and NIP jointly on the same shared backbone:
\begin{equation}
  \mathcal{L} = \mathcal{L}_{\mathrm{IRL}} + \lambda_t \mathcal{L}_{\mathrm{NIP}}.
\end{equation}

\noindent\textbf{Warm-up schedule.} Training proceeds in two phases. During the first $10\%$ of steps, only $\mathcal{L}_{\mathrm{IRL}}$ is active ($\lambda_t = 0$), giving the item space its initial co-engagement structure. After warm-up, $\lambda_t$ ramps linearly from $0.1$ to $1.0$.

\noindent\textbf{Joint training dynamics.} IRL keeps the backbone anchored to the item-pair structure while NIP trains user histories to retrieve future items. As a result, the indexed item space and the query side evolve together, rather than forcing the query side to align after the item space has been frozen. Section~\ref{sec:exp-joint-vs-seq} compares this configuration with alternatives that ablate task coupling, the stop-gradient mechanism, and backbone sharing.

\subsection{Serving Architecture}
\noindent\textbf{Offline indexing.} EGR separates item-index construction from query generation. Item embeddings can be computed periodically or in near real-time, depending on the cost and freshness tradeoff, and are written to an ANN index; because embeddings are L2-normalized, maximum-inner-product search is equivalent to cosine retrieval.

\noindent\textbf{Query retrieval.} EGR supports two query-serving modes. In our deployed pipeline (Section~\ref{sec:exp-online}), each user's query embedding is generated once per day through the user-history path and stored in the serving layer. For each request, the system loads this precomputed embedding, performs online ANN search against the current product index, and forwards the top-$K$ products to downstream ranking. For surfaces where intent changes within a session, the same encoding path can run at request time: the system serializes the user's recent item engagements, runs the user-history path online, retrieves top-$K$ items from the index, and forwards candidates to downstream ranking at a higher per-request serving cost.

\noindent\textbf{Practical advantages.} This architecture has three operational advantages: (i) a richer item encoder adds no per-request cost when items are indexed offline; (ii) the output is a dense vector, so serving can use standard ANN infrastructure rather than a token-generation stack; and (iii) new or updated items can be re-encoded and inserted into the index without rebuilding a semantic-ID vocabulary or retraining a generator over item-specific tokens.

\section{Experiments}
\label{sec:experiments}

\subsection{Experimental Setup}
\label{sec:exp-setup}

\noindent\textbf{Datasets.} For public-benchmark evaluation, we use the Amazon Reviews datasets~\cite{mcauley2015amazon} on three categories: Beauty, Sports, and Toys. Following~\cite{rajput2023tiger,liu2025onerecthink}, we apply 5-core filtering and use leave-one-out splitting. For industrial evaluation, we use Snap Dynamic Product Ads (DPA) logs of user-product engagement events (we use ``product'' and ``item'' interchangeably throughout this section), and hold out 1M user sequences for evaluation, using the remaining sequences for training.

\noindent\textbf{Metrics.} We report Recall@$10$ (R@10) for offline retrieval. For the online deployment, we report relative lifts in impressions, click-through rate (CTR), and conversion rate (CVR). Per Snap's policy, experiments on Snap data are reported only as relative numbers.

\noindent\textbf{Inputs.} Both paths use the model's chat template. The product path pairs the system instruction ``Represent this product for retrieval.'' with a user message of the form \texttt{Title: \ldots, Brand: \ldots, Category: \ldots}. The user-history path pairs the system instruction ``Predict the next product this user will interact with.'' with a message that lists each engagement as $\langle\text{conv\_token}\rangle$ followed by product text, with events joined by `|' separators. For example: \texttt{<view> Title: Galaxy Watch, Brand: Samsung, Category: Electronics | <add\_to\_cart> Title: Coffee Maker, Brand: Philips, Category: Kitchen | <purchase> Title: Running Shoes, Brand: Nike, Category: Footwear}.

\noindent\textbf{Implementation.} We use Qwen3-VL-Embedding 2B~\cite{qwen3vlembedding} as the default backbone with last-token pooling, and add a two-layer MLP projection head (LayerNorm, GELU, dropout) trained from scratch to produce L2-normalized $512$-dimensional embeddings. LoRA~\cite{hu2022lora} adapters use $r{=}128$, $\alpha{=}256$ on the q/k/v/o attention projections, yielding approximately $41$M trainable parameters, mostly in the LoRA matrices and the projection head. We optimize the LoRA matrices, projection head, and learned temperature with AdamW at learning rate $2{\times}10^{-5}$, using a short learning-rate warm-up followed by cosine decay. The canonical recipe trains for one epoch on $8$ H100 GPUs with per-GPU batch size $64$ and maximum history length $20$, corresponding to approximately $460$ input tokens on average. In-batch negatives are gathered across all Distributed Data Parallel (DDP) ranks, giving $B \times W$ negatives per anchor for local batch size $B$ and $W$ GPUs.

\subsection{Public Amazon Reviews Benchmark}
\label{sec:exp-amazon}

We evaluate EGR on three public Amazon Reviews categories: Beauty, Sports, and Toys, and compare against representative sequential recommendation baselines. Table~\ref{tab:amazon-public} reports R@10 on each category.

\begin{itemize}
  \item \textbf{BERT4Rec~\cite{sun2019bert4rec}.} A bidirectional transformer trained with a masked-item (cloze) objective that predicts items via softmax over the item vocabulary.
  \item \textbf{SASRec~\cite{kang2018sasrec}.} A causal self-attention sequence model trained for next-item prediction with learned item embeddings as model parameters.
  \item \textbf{TIGER~\cite{rajput2023tiger}.} An SID-based generative retriever that quantizes item content embeddings into discrete semantic identifier sequences and trains a transformer to generate them autoregressively.
  \item \textbf{HSTU~\cite{zhai2024hstu}.} A hierarchical sequential transducer for generative recommendation at industrial scale.
\end{itemize}

\begin{table}[t]
  \caption{Recall@10 on Amazon Reviews. Best per column in \textbf{bold}.}
  \Description{Table reporting Recall@10 on three Amazon Reviews categories (Beauty, Sports, Toys) across five methods: BERT4Rec, SASRec, TIGER, HSTU, and EGR. EGR achieves the highest Recall@10 in every category.}
  \label{tab:amazon-public}
  \small
  \setlength{\tabcolsep}{8pt}
  \begin{tabular}{@{}lccc@{}}
    \toprule
    Method & Beauty & Sports & Toys \\
    \midrule
    BERT4Rec~\cite{sun2019bert4rec} & 0.0396 & 0.0175 & 0.0332 \\
    SASRec~\cite{kang2018sasrec}    & 0.0607 & 0.0301 & 0.0626 \\
    TIGER~\cite{rajput2023tiger}    & 0.0623 & 0.0347 & 0.0547 \\
    HSTU~\cite{zhai2024hstu}        & 0.0652 & 0.0343 & 0.0566 \\
    \textbf{EGR (ours)}    & \textbf{0.0655} & \textbf{0.0362} & \textbf{0.0807} \\
    \bottomrule
  \end{tabular}
\end{table}

We find that EGR outperforms the baseline methods on all three categories: it is essentially tied with HSTU on Beauty ($0.0655$ vs.\ $0.0652$), improves over TIGER on Sports ($0.0362$ vs.\ $0.0347$), and obtains a larger margin over the next-best baseline on Toys ($0.0807$ vs.\ $0.0626$).

\subsection{Snap DPA Offline Results}
\label{sec:exp-offline}

\subsubsection{\textbf{Data Scaling}}
\label{sec:exp-data-scaling}

\begin{figure}[b]
  \centering
  \begin{tikzpicture}
    \begin{semilogxaxis}[
      width=\linewidth, height=5.2cm,
      xlabel={Training users},
      ylabel={Relative recall},
      xtick={10000, 100000, 1000000, 10000000},
      xticklabels={10K, 100K, 1M, 10M},
      xminorticks=false,
      yminorticks=false,
      legend pos=south east,
      legend style={
        font=\scriptsize, draw=none, fill=none, row sep=-2pt,
      },
      legend cell align=left,
      grid=major, grid style={dashed, gray!25},
      ymin=0.95, ymax=1.35,
      ytick={1.0, 1.1, 1.2, 1.3, 1.4, 1.5},
      mark size=2.2pt,
      tick label style={font=\scriptsize},
      label style={font=\small},
    ]

      \addplot+[mark=square*, thick] coordinates {
        (10000, 1.000) (100000, 1.114) (1000000, 1.183) (10000000, 1.270)
      };
      \addlegendentry{R@10}
    \end{semilogxaxis}
  \end{tikzpicture}
  \caption{Data scaling of EGR. R@10 normalized to its value at $10$K training users.}
  \Description{Line chart showing relative Recall@10 as training users increase from 10K to 10M. R@10 is normalized to its value at 10K training and rises from 1.00 to 1.27.}
  \label{fig:data-scaling}
\end{figure}
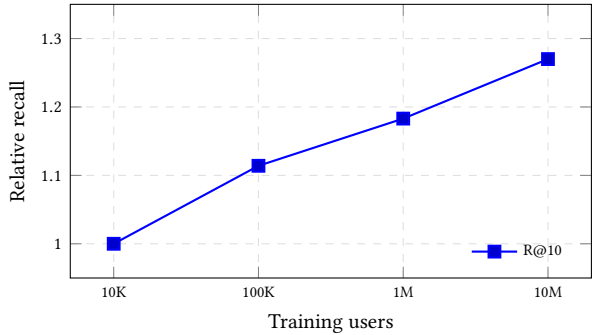

We train EGR on four matched-event subsets ranging from $10$K to $10$M users. Figure~\ref{fig:data-scaling} reports R@10, normalized to its value at $10$K training users.

\noindent\textbf{Findings.} R@10 increases monotonically with training data, growing by $27\%$ from $10$K to $10$M users with no observed saturation. More training data continues to improve retrieval quality at industrial scale.

\subsubsection{\textbf{Cold-Start and Frequency Strata}}
\label{sec:exp-stratified}

We stratify the offline test set by the training-input frequency of each target product, defined as the number of times the target product appeared in the training data. The cold-start stratum (frequency $0$) contains products the model never saw as training inputs; the head stratum ($>15$) contains the most frequently seen products. Table~\ref{tab:stratified-eval} reports relative R@10, normalized so the head bucket equals $100\%$.


\begin{table}[t]
  \caption{EGR retrieval, stratified by each target product's training-input frequency. Count $=0$: the target never appeared in training input. R@10 normalized to the head bucket ($\text{freq}>15$) at $100\%$ (higher is better).}
  \Description{Table stratifying retrieval performance by the target product's training-input frequency across nine buckets from 0 (cold start) to greater than 15 (head). R@10 is reported relative to the head bucket. The cold-start bucket reaches 94.9\%; mid-frequency buckets reach the highest values.}
  \label{tab:stratified-eval}
  \small
  \setlength{\tabcolsep}{6pt}
  \begin{tabular}{@{}lcc@{}}
    \toprule
    Count & \% of test samples & rel R@10 \\
    \midrule
    0      & $31.3\%$ & $94.9\%$  \\
    1      & $8.0\%$  & $99.1\%$  \\
    2      & $6.0\%$  & $111.5\%$ \\
    3      & $4.3\%$  & $112.7\%$ \\
    4      & $3.6\%$  & $117.6\%$ \\
    5      & $2.8\%$  & $115.4\%$ \\
    6--10  & $9.4\%$  & $119.1\%$ \\
    11--15 & $5.3\%$  & $116.8\%$ \\
    $>15$  & $29.2\%$ & $100.0\%$ \\
    \bottomrule
  \end{tabular}
\end{table}

\noindent\textbf{Findings.} EGR retrieves never-seen products with limited degradation: cold-start R@10 trails the head bucket by only $5.1\%$. Mid-frequency strata ($2$--$15$) score highest. This non-monotonic pattern suggests that factors beyond target frequency affect retrieval quality. Because the buckets are defined only by target frequency, this experiment does not isolate those factors. Thus, the strong cold-start result indicates robustness to unseen products but not parity with the best-performing mid-frequency strata.

\subsubsection{\textbf{Multimodal Input}}
\label{sec:exp-multimodal}

The default Qwen3-VL-Embedding 2B backbone supports multimodal product input through its vision tower. We compare two configurations on the same backbone: \emph{text-only}, which encodes each product using title, brand, and category, and \emph{multimodal}, which additionally feeds the product image through the vision tower. Table~\ref{tab:multimodal} reports R@10 normalized to the text-only configuration at $100\%$.


\begin{table}[t]
  \caption{Multimodal vs.\ text-only product input on Qwen3-VL-Embedding 2B. R@10 normalized to text-only at $100\%$ (higher is better).}
  \Description{Table comparing text-only and text-plus-image product input on the Qwen3-VL-Embedding 2B backbone. R@10 is normalized to text-only at 100\%; adding the product image lifts R@10 to 106.9\%.}
  \label{tab:multimodal}
  \small
  \setlength{\tabcolsep}{8pt}
  \begin{tabular}{@{}lc@{}}
    \toprule
    Modality & rel R@10 \\
    \midrule
    Text-only       & $100.0\%$ \\
    Text$+$image    & \textbf{$106.9\%$} \\
    \bottomrule
  \end{tabular}
\end{table}

\noindent\textbf{Findings.} Adding the product image improves R@10 by $+6.9\%$ over text-only on the same backbone. EGR can therefore use multimodal product evidence without changing the retrieval architecture.

\subsection{Snap DPA Ablation Study}
We study the joint-training strategy and the NIP loss-weight schedule.

\subsubsection{\textbf{Effective Joint Training}}\label{sec:exp-joint-vs-seq}

We compare EGR against four alternative configurations. Table~\ref{tab:joint-vs-others} reports R@10 normalized so the full EGR recipe equals $100\%$.

\begin{itemize}
  \item \textbf{EGR (ours).} The shared backbone is trained on IRL and NIP jointly, with stop-gradient on the target embedding in NIP.
  \item \textbf{EGR w/o stop-grad.} Same as EGR (ours) but without the stop-gradient in NIP.
  \item \textbf{NIP only.} The same backbone is trained on NIP alone.
  \item \textbf{IRL only.} The same backbone is trained on IRL alone.
  \item \textbf{Two models.} Model 1 is trained on IRL and frozen; Model 2 is trained on NIP using embeddings produced by Model 1 as input.
\end{itemize}


\begin{table}[t]
  \caption{Joint-training ablation. R@10 normalized to the full EGR recipe at $100\%$ (higher is better). JT: IRL and NIP optimized jointly in a shared backbone. SG: stop-gradient on the target embedding in NIP.}
  \Description{Table comparing five training configurations on R@10 relative to the full EGR recipe. EGR with joint training and stop-gradient scores 100\%, EGR without stop-gradient 95.3\%, NIP-only 93.7\%, IRL-only 26.6\%, and the two-model separate-training baseline 86.2\%.}
  \label{tab:joint-vs-others}
  \footnotesize
  \setlength{\tabcolsep}{6pt}
  \begin{tabular}{@{}lccccc@{}}
    \toprule
    Method & IRL & NIP & JT & SG & rel R@10 \\
    \midrule
    EGR (ours)        & \checkmark & \checkmark & \checkmark & \checkmark & \textbf{$100.0\%$} \\
    EGR w/o stop-grad & \checkmark & \checkmark & \checkmark &            & $95.3\%$ \\
    NIP only          &            & \checkmark &            & ---        & $93.7\%$ \\
    IRL only          & \checkmark &            &            & ---        & $26.6\%$ \\
    \midrule
    Two models        & \checkmark & \checkmark &            & ---        & $86.2\%$ \\
    \bottomrule
  \end{tabular}
\end{table}

\noindent\textbf{Findings.} Table~\ref{tab:joint-vs-others} shows four effects. First, removing IRL while keeping NIP reduces relative R@10 from $100.0\%$ to $93.7\%$, indicating that item-pair contrastive learning remains useful even when the history-to-target loss is active. Second, IRL alone reaches only $26.6\%$, so co-engagement structure is a useful item-space anchor but is not sufficient for next-item retrieval. Third, removing the stop-gradient on the NIP target branch lowers R@10 to $95.3\%$, consistent with the stop-gradient stabilizing the indexed item space. Fourth, freezing the item side and training a separate query model reaches only $86.2\%$. Jointly shaping the item and query spaces, with stop-gradient on the target branch, is more effective than ablating either objective or aligning a query generator to a frozen item encoder.

\subsubsection{\textbf{Loss Schedule}}
\label{sec:exp-loss-schedule}

We compare three loss-schedule variants for the joint training objective. \emph{Joint} (ours): the IRL weight stays at $1$ throughout, while the NIP weight $\lambda_t$ ramps linearly from $0.1$ to $1.0$ after warm-up. \emph{Balanced}: IRL weight decays as $1{-}\lambda_t$ while NIP weight ramps as $\lambda_t$, so total loss weight stays constant. \emph{Sequential}: IRL is active only during the first half of training; NIP is active only during the second half, with no overlap phase. Table~\ref{tab:loss-schedule} reports R@10 normalized to Joint at $100\%$.

\begin{table}[t]
  \caption{Ablation of training loss schedules. Results are normalized Recall@10 relative to the Joint schedule.}
  \Description{Comparison of different IRL and NIP training schedules measured by normalized Recall@10.}
  \label{tab:loss-schedule}
  \small
  \centering
  \setlength{\tabcolsep}{6pt}
  \renewcommand{\arraystretch}{1.1}
  \begin{tabular}{@{}lccc@{}}
    \toprule
    Variant & IRL Schedule & NIP Schedule & rel R@10 \\
    \midrule
    Joint (ours)
      & fixed at $1$
      & $\lambda_t$: $0.1 \rightarrow 1.0$
      & 100.0\% \\

    Balanced
      & $1 - \lambda_t$
      & $\lambda_t$
      & 97.9\% \\

    Sequential
      & first 50\% only
      & last 50\% only
      & 74.4\% \\
    \bottomrule
  \end{tabular}
\end{table}

\noindent\textbf{Findings.} Maintaining IRL throughout joint training is important. \emph{Balanced} trails \emph{Joint} by only about $2\%$, suggesting that decayed IRL still preserves enough product-space supervision for the indexed space to remain mostly intact. \emph{Sequential} trails by about $26\%$: when IRL is turned off after the first half of training, NIP must train against a product space that is no longer directly anchored by co-engagement structure. The overlap phase in which both losses are active is therefore central to the training recipe.

\subsection{Online A/B Test}
\label{sec:exp-online}

We conducted a two-week online A/B test in Snap DPA. The treatment arm included EGR as an additional retrieval source while holding the total candidate quota fixed; the control arm retained the existing production retrieval stack without EGR. Each arm received $10\%$ of traffic through Snap's experimentation platform. EGR operated as a hybrid batch-and-online retrieval pipeline: each user's query embedding was computed once per day and stored for serving; at request time, the system used this embedding to perform real-time ANN search and forward the retrieved candidates to downstream ranking. Downstream ranking, auction, policy, and rendering were held fixed across arms. We report relative lifts over the control arm on Snap's standard DPA engagement-quality metrics (impressions, CTR, CVR); absolute values are withheld per Snap's policy.

\begin{table}[t]
  \caption{Online A/B test in Snap DPA over two weeks ($10\%$ / $10\%$ user traffic split). Relative lifts of the EGR-enabled treatment arm over the control arm; absolute values are withheld per Snap's policy.}
  \Description{Table reporting relative lifts of the EGR-enabled treatment arm over control in an online A/B test: impressions +0.15\%, CTR +0.23\%, CVR +2.91\%.}
  \label{tab:online-ab}
  \small
  \setlength{\tabcolsep}{8pt}
  \begin{tabular}{@{}lc@{}}
    \toprule
    Metric & Relative lift \\
    \midrule
    Impressions & $+0.15\%$ \\
    CTR         & $+0.23\%$ \\
    CVR         & $+2.91\%$ \\
    \bottomrule
  \end{tabular}
\end{table}

\noindent\textbf{Findings.} Adding EGR to the production retrieval stack increases impressions by $+0.15\%$, CTR by $+0.23\%$, and CVR by $+2.91\%$ (Table~\ref{tab:online-ab}). Because the total candidate quota is held constant across arms, the CTR and CVR gains reflect improved retrieval quality rather than additional candidate allocation. Of all final impressions delivered to users, $5.3\%$ come from EGR. At the source level, EGR's CTR and CVR are $1.55\times$ and $1.53\times$ the respective averages across all production retrieval sources, including user-to-item (U2I), item-to-item (I2I), and non-personalized retrievers. After the A/B test, EGR was rolled out across DPA and now serves $100\%$ of users.

\section{Conclusion}
We presented EGR, an embedding-native framework for generative retrieval. Rather than generating discrete item identifiers or training separate item and query encoders, EGR uses a single shared LLM backbone to produce both indexed item embeddings and user-conditioned query embeddings. The model is trained jointly with a symmetric item-pair contrastive loss and a stop-gradient history-to-target loss, while retaining the operational simplicity of standard ANN serving.

Experimental results demonstrate that EGR outperforms prior sequential recommendation and generative retrieval baselines on Amazon Reviews. On Snap DPA, EGR scales monotonically with training data, remains strong on cold-start items, and benefits from multimodal input. In a live production deployment, incorporating EGR into the retrieval stack yields a $+2.91\%$ conversion-rate improvement. Together, these results show that embedding-native generative retrieval can simplify system design while improving both offline retrieval quality and online advertising performance.

\bibliographystyle{ACM-Reference-Format}
\bibliography{main}

\end{document}